\documentstyle[11pt,aaspp4]{article}

\def\be{\begin{equation}}
\def\ee{\end{equation}}
\def\go{\mathrel{\raise.3ex\hbox{$>$}\mkern-14mu
             \lower0.6ex\hbox{$\sim$}}}
\def\lo{\mathrel{\raise.3ex\hbox{$<$}\mkern-14mu
             \lower0.6ex\hbox{$\sim$}}}

\def\nuqpo{\nu_{\rm QPO}}
\def\nuk{\nu_K}
\def\rin{r_{\rm in}}
\def\nus{\nu_s}
\def\mdot{\dot M}
\def\gps{{\rm g\,s^{-1}}}
\def\Cdot{\dot C}
\def\numax{\nu_{\rm max}}
\def\msun{M_\odot}
\def\hz{{\rm Hz}}

\begin{document}

\title{Kilohertz Quasi-Periodic Oscillations, Magnetic Fields and Mass of 
Neutron Stars in Low-Mass X-Ray Binaries}

\author{Dong Lai, Richard Lovelace and Ira Wasserman}
\affil{Center for Radiophysics and Space Research, Department of
Astronomy, Cornell University,
Ithaca, NY 14853\\
(submitted to ApJ, March 11 1999)}

\begin{abstract}
It has recently been suggested that the maximum observed quasi-periodic
oscillation (QPO) frequencies, $\numax$, 
for several low-mass X-ray binaries, particularly
4U 1820-30, correspond to the orbital frequency at the inner-most stable orbit
of the accretion disk. This would imply that the neutron stars in these
systems have masses $\go 2~M_\odot$, considerably larger than 
any well-measured neutron star mass.
We suggest that the levelling off of $\nuqpo$ 
may be also understood in terms of a steepening
magnetic field which, although possibly dipolar at the stellar surface,
is altered substantially by disk accretion, and presents a ``wall'' to the
accretion flow that may be outside the innermost stable orbit. General
relativistic effects add to the flattening of the $\nuqpo-\mdot$ relation
at frequencies below the Kepler frequency at the innermost stable orbit.
We offer two other possible ways to reconcile the low value of
$\numax$ ($\approx 1060$ Hz for 4U 1820-30)
with a moderate neutron star mass, $\approx 1.4\msun$: at sufficiently
large $\mdot$, either (i) the disk terminates in a very thin boundary layer
near the neutron star surface, or (ii) $\nuqpo$ is not the orbital frequency
right at the inner edge of the disk, but rather at a somewhat larger radius,
where the emissivity of the disk peaks.

\end{abstract}


\section{Introduction}

Recent observations with Rossi X-ray Timing Explorer (RXTE) have revealed 
kilohertz quasi-periodic oscillations (QPOs) in at least eighteen 
low-mass X-ray binaries (LMXBs; see Van der Klis 1998a,b for a review; 
also see Eric Ford's QPO web page at 
http://www.astro.uva.nl/ecford/qpos.html for updated information).  
These kHz QPOs are characterized by 
their high levels of coherence (with $\nu/\Delta\nu$ up to $100$), 
large rms amplitudes (up to $20\%$), and wide span of frequencies 
($500-1200$ Hz). In almost all sources, 
the X-ray power spectra show twin kHz peaks moving up and down in frequency
together as a function of photon count rate, with the separation frequency
roughly constant (The clear exceptions are Sco X-1 and 4U 1608-52, 
van der Klis et al.~1997, Mendez et al.~1998a; see also
Psaltis et al.~1998. In Aql X-1, only a single QPO has been detected.). 
Moreover, in several sources,
a third, nearly coherent QPO has been detected
during one or more X-ray bursts, at a frequency
approximately equal to the frequency difference between the 
twin peaks or twice that value. (An exception is 4U 1636-53, 
Mendez et al.~1998b.) 
The observations suggest a generic beat-frequency model 
where the QPO with the higher frequency is associated
with the orbital motion at some preferred orbital radius
around the neutron star, while the lower-frequency QPO results
from the beat between the Kepler frequency and the neutron star 
spin frequency. It has been suggested that this preferred radius
is the magnetosphere radius (Strohmayer et al.~1996) or the sonic
radius of the disk accretion flow (Miller, Lamb and Psaltis 1998; see also
Klu\'zniak et al.~1990).
The recent observational findings (e.g., the variable frequency 
separations for Sco X-1 and 4U 1608-52) indicate that the ``beat'' is 
not perfect, so  perhaps a boundary 
layer with varying angular frequencies, rather than simply the neutron
star spin, is involved.

This paper is motivated by recent RXTE observation of the bright globular
cluster source 4U 1820-30 (Zhang et al.~1998), which has 
revealed that, as a function of X-ray photon count rate, $\Cdot$,
the twin QPO frequencies increase 
roughly linearly for small photon count rates 
($\Cdot\approx
1600-2500$~cps)
and become independent of $\Cdot$ for larger photon count
rates 
($\Cdot\approx 2500-3200$~cps). 
(The QPOs become unobservable for still higher
count rates.) It was suggested that the $\Cdot-{\rm independent}$
maximum frequency
($\numax=1060\pm 20$~Hz) of the upper QPO corresponds to the orbital frequency 
of the disk at the {\it inner-most stable circular orbit} (ISO) 
as predicted by general relativity. 
This would imply that the NS has mass of $2.2M_\odot$ 
(assuming a spin frequency of $275$~Hz). It has also been noted earlier
(Zhang et al.~1997), based on the narrow range of the maximal 
QPO frequencies ($\numax\approx 1100-1200$ Hz) in at least six sources
(which have very different X-ray luminosities), that 
these maximum frequencies correspond to the Kepler
frequency at the ISO, which then implies that the 
neutron star masses are near $2M_\odot$ (see also Kaaret et al.~1997).

The neutron star masses inferred from identifiying $\numax$ with
the Kepler frequency at the ISO would, if confirmed,
be of great importance for constraining the properties of
neutron stars and for understanding the recycling processes 
leading to the formation of millisecond pulsars.
However, while it is tempting to identify
$\numax$ with the
the orbital frequency at the ISO, this 
seemingly natural
interpretation may not be true.
One clue that this identification may not be correct is that
the inferred neutron star masses are substantially above the
masses of those neutron stars for which accurate determinations
are available (Thorsett \& Chakrabarty 1999) even though spin-up to 
$\nus\sim 300\,\hz$ only requires accretion of a very small
amount of material ($\ll\msun$; \S 2).
The cause of the flattening of the $\nuqpo-\Cdot$ correlation, and
the value of the maximum frequency, are still not understood
(and the existence of a plateau in $\nuqpo$ with
increasing $\mdot$ is debatable; e.g. Mendez et al.~1998c).
We suggest in \S 3 that the steepening of the magnetic 
field, expected near the accreting neutron star, together with general
relativistic effect, naturally leads to the flattening in the $\nu_{\rm QPO}$
-$\dot M$ correlation. In \S 4 we advocate two alternative interpretations
of the maximum QPO frequency without invoking excessively large neutron star
masses. 


\section{Possible Problems with Neutron Star Masses $\go 2~M_\odot$}

The most important concern for the inferred neutron star mass of
$\go 2M_\odot$ is an empirical one. LMXBs have long been thought 
(e.g., Alpar et al.~1982) to be
the progenitors of binary millisecond radio pulsars. The recent
discovery of binary X-ray pulsar SAX J1808-3658 (with spin period 2.5~ms and
orbital period 2~hrs; Wijnands \& van der Klis 1998; Chakrabarty \& 
Morgan 1998) appears to confirm this link. 
Measurements of neutron star masses
in radio pulsar binaries give values in a narrow range
around $M\simeq 1.4\msun$; the data are consistent with a neutron star
mass function that is flat between $\go 1.1\msun$ and $\lo 
1.6\msun$ at 95\% CL (Thorsett \& Chakrabarty 1999, Finn 1994).
The masses of neutron stars in X-ray binaries are also consistent
with $M\simeq 1.4\,M_\odot$ (e.g., van Kerkwijk et al.~1995). 
Of particular interest is the $5.4$ ms recycled
pulsar B1855+09 with a white dwarf companion: this system
is thought to have gone through a LMXB phase (Phinney \& Kulkarni 1994),
and contains a neutron star with 
$M=1.41\pm 0.10M_\odot$ 
(Thorsett \& Chakrabarty 1999; earlier Kaspi et al. 1994
estimated $M=1.50\pm^{0.26}_{0.14}M_\odot$). 
The 23 ms pulsar PSR B1802-07, which is in a white dwarf
binary that is also thought to have gone through the LMXB
phase, has an inferred mass
$M=1.26\pm^{0.15}_{0.67}M_\odot$ ($95\%$ confidence;
Thorsett \& Chakrabarty 1999). 

If $1.4~M_\odot$ is the mass of the neutron star immediately after 
its formation in core collapse, then to make a $2.2\,M_\odot$ 
object would require accretion of material of at least $0.8\,M_\odot$.
Such large accretion mass may be problematic. 
If we neglect torques on the star due to the interaction of its magnetic
field and the accretion disk, the added mass
needed to spin up the NS to a spin frequency $\nu_s=\Omega_s/(2\pi)$ is
\be
\Delta M\simeq {I\Omega_s\over\sqrt{GMr_{\rm in}}}
\simeq 0.07\msun {I_{45}\over\sqrt{M_{1.4}r_6}}
\left({\nu_s\over 300\,{\rm Hz}}\right),
\label{eq:delm1}\ee
where $I=10^{45}I_{45}$~g~cm$^2$ is the moment of inertia, 
$M=1.4M_{1.4}M_\odot$ is the neutron star mass and
and $r_{\rm in}=10r_6$~km is the radius of the inner edge of the accretion
disk, which could correspond to either the stellar surface (radius $R$)
or the inner-most stable orbit (ISO) in the absence of a magnetic field
strong enough to influence the flow substantially
(see Cook et al.~1994).
When the neutron star magnetic field is strong enough, the inner radius
$\rin$ corresponds to the Alfv\'en radius. (We note that the positions of
all known millisecond pulsars and binary pulsars in the $P-\dot P$
diagram for radiopulsars are consistent with spinup via accretion onto
neutron stars with dipolar surface fields $\go 10^{8-9}$ G.)
For magnetic accretion, we expect
\be
I\dot\Omega_s=\dot M\sqrt{GMr_{\rm in}} f(\omega_s),
\ee
where $\omega_s=\Omega_s/\Omega_K(r_{\rm in})$, with $\Omega_K(r_{\rm in})$
the Kepler frequency at $r_{\rm in}$. The dimensionless
function $f(\omega_s)$ includes 
contributions to the angular momentum transport from magnetic stresses and 
accreting material. It is equal to zero at some 
equilibrium $\omega_s$, but the actual form of $f(\omega_s)$
depends on details of the magnetic field -- disk interaction. 
Treating $r_{\rm in}$ as a constant, we find 
\begin{eqnarray}
\Delta M &=& 
{I\Omega_s\over\sqrt{GMr_{\rm in}}}\left[{1\over\omega_s}
\int_0^{\omega_s}\!\!{d\omega_s'\over f(\omega_s')}\right]\nonumber\\
&=&0.07\msun{I_{45}\over\sqrt{M_{1.4}r_6}}\biggl[\omega_s^{-1}
\ln\biggl({1\over 1-\omega_s}\biggr)\biggr]\biggl({\nu_s\over 
300\,\hz}\biggr)\nonumber\\
&=& 0.04M_\odot I_{45}M_{1.4}^{-2/3}\left[\omega_s^{-4/3}
\ln\left({1\over 1-\omega_s}\right)\right]\left({\nu_s\over 300~{\rm Hz}}\right)^{4/3},
\label{eq:delm2}
\end{eqnarray}
where, in the last two lines, we have adopted a simple functional form 
$f(\omega_s)=1-\omega_s$; generically,
\be
\Delta M\approx 0.07\msun{I_{45}\over\sqrt{M_{1.4}r_6}}
\biggl({\nu_s\over300\,\hz}\biggr)\psi(\omega_s),
\ee
where $\psi(\omega_s)\to 1$ for $\omega_s\ll\omega_{s,c}$, assuming that the torque
tends to zero at a critical value $\omega_s=\omega_{s,c}$. 

Large $\Delta M$ is possible
if there is a lengthy phase of accretion with nearly zero net torque (e.g. accreting
$0.8\msun$ at a mean accretion rate of $\mdot=10^{17}\gps$ would require about
400 Myr) following a much shorter phase of spin-up to $\omega_s\to\omega_{s,c}$
(e.g. accreting $0.05\msun$ at $\mdot=10^{17}\gps$ would require 30 Myr).
If magnetic field decays during accretion (e.g. Taam \& van den Heuvel 1986,
Shibazaki et al. 1989), then the spin-up phase would have been even shorter.
(Spin diffusion due to alternating or stochastic epsiodes of spin-up and spin-down
[e.g. Bildsten et al. 1997, Nelson et al. 1997] might be allowed -- but constrained
-- in such a picture.) To accomodate masses as large as $2\msun$, these LMXBs must
be rather old and must have spun up rapidly at first, and then not at all for
$\go 90\%$ of their lifetimes. Gravitational radiation might provide a mechanism
for enforcing virtually zero net torque during the bulk of accretion (Bildsten
1998, Andersson et al. 1999). But equations (\ref{eq:delm1}) and (\ref{eq:delm2})
show that only very small $\Delta M$ is {\it required} to achieve $\nus\sim
300\,\hz$, irrespective of the mechanism responsible for halting spin-up at
such frequencies.


\section{Steepening Magnetic Fields Near the Accreting Neutron Star}

We shall adopt, as a working hypothesis, that the upper QPO 
frequency is approximately equal to the Kepler frequency at a
certain critical radius of the disk (Strohmayer et al.~1996; 
Miller, Lamb \& Psaltis 1998; van der Klis 1998)\footnote{In the model 
of Titarchuk et al.~(1998) , the QPO corresponds of vertical oscillation of 
the disk boundary layer, but the oscillation frequency is equal to the local 
Kepler frequency. Even in the ``non-beat'' frequency model of Stella and 
Vietri (1998), the upper QPO frequency still corresponds to the orbital
frequency.} that is determined by 
the combined effects of general relativity and stellar magnetic field.
For sufficiently strong magnetic fields, the disk may be
truncated near this radius, where matter flows out of the disk and is
funneled toward the neutron star. This critical radius then corresponds to the
usual Alfv\'en radius (Strohmayer et al.~1996). Even if the fields are relatively
weak ($10^7-10^8$~G) and the field geometry is such that 
matter remains in the disk, the magnetic stress can still 
slow down the orbital motion in the inner disk by taking away 
angular momentum from the flow, and accreting gas then plunges toward
the star at supersonic speed -- a process that is also accelerated 
by relativistic instability. In this case, the critical radius
would correspond to the sonic point of the flow (Lai 1998). 
We neglect the possible role of radiative forces discussed by 
Miller et al.~(1998). As emphasized by van der Klis (1998a), 
the fact that similar QPO frequencies ($500-1200$~Hz) 
are observed in sources with vastly different average luminosities
(from a few times $10^{-3}L_{\rm Edd}$ to near $L_{\rm Edd}$)
suggests that radiative effects cannot be the only factor 
that induces the correlation of the QPO frequency and the X-ray flux
for an individual source. 

Despite many decades of theoretical studies (e.g., 
Pringle \& Rees 1972; Lamb et al.~1973; Ghosh \& Lamb 1979; 
Arons 1987; Spruit \& Taam 1990; Aly 1991; Sturrock 1991;
Shu et al.~1994; 
Lovelace, Romanova \& Bisnovatyi-Kogan ~1995, ~1999; Miller \& Stone 1997), there remain considerable
uncertainties on the nature of the stellar magnetic field -- disk interactions.
Among the issues that are understood poorly are the
transport of magnetic field in the disk, 
the configuration of the field threading the disk,
and the nature of outflows from the disk.
To sidestep these complicated questions,
we adopt a simple phenomenological prescription
for the vertical and azimuthal components of
the magnetic field on the disk,
\be
B_z=B_0\left({R\over r}\right)^n,~~~~
B_\phi=-\beta B_z,
\label{eq:Bfield}
\ee
where $B_\phi$ is evaluated at the upper surface of the
disk, and $\beta$ is the
azimuthal pitch angle of the field.  
If we neglect the GR effect, the critical radius $r_{\rm in}$ is located 
where the magnetic field stress dominates the angular momentum transport in the
disk, and it is approximately given by the condition
\be
\dot M {d\sqrt{GMr}\over dr}=-r^2B_zB_\phi;
\label{eq:balance}\ee 
using the {\it ansatz} equation (\ref{eq:Bfield}), and assuming Keplerian
rotation (which may break down in a boundary layer near $\rin$; e.g. Lovelace
et al. 1995), we find
\be
r_{\rm in}=R\left(2\beta{B_0^2R^3\over\dot M\sqrt{GMR}}\right)^{2/(4n-5)}.
\ee
and the Kepler frequency at $\rin$ is 
\be
\nu_K(r_{\rm in})\propto \dot M^{3/(4n-5)}.
\label{eq:scale}\ee
For a ``dipolar'' field configuration, $n=3$ and $\nuk(\rin)\propto
\mdot^{3/7}$, as is well-known, but for smaller values of $n$, the
dependence steepens; for example, $\nuk(\rin)\propto\mdot$ for a
``monopole'' field, $n=2$. The observed correlation $\nuqpo\propto\Cdot$
may require $n<3$, although the relationship between $\Cdot$ and $\mdot$
is unclear (Mendez et al. 1998c).

Unusual field topologies are possible as the disk approaches the surface
of the neutron star. Values of $n\neq 3$ (and even violation of power-law
scaling) might occur naturally, for open field configurations, which may be 
prevalent because of differential rotation between the star and the disk
(e.g. Lovelace et al. 1995). MHD winds driven off a disk could
also result in $n\neq 3$ (e.g. Lovelace et al.
1995, Blandford \& Payne 1982). Disks that are fully 
(Aly 1980, Riffert 1980) or partially (Arons 1993) diamagnetic will also
have non-dipolar variation in field strength near their inner edges
(see also \S 3.1 below). None of these possibilities {\it requires}
the field to be substantially non-dipolar {\it at the stellar surface},
although for disks that penetrate close to the star (at $\rin-R\lo R$)
any non-dipolar field components, if strong enough, would be significant.

A particular field configuration that could explain the observed variation
of $\nuqpo$ with $\Cdot$ might have $n<3$ at moderate values of $\rin$, leading
to a strong correlation between $\nuqpo$ and $\mdot$ (and hence $\Cdot$).
As $\mdot$ rises, the disk approaches the star, and the field topology
could become more complex, resulting in additional, non-power-law radial
steepening of the field strength. As is argued below,
this could happen even if the field is dipolar at the surface of the star, 
particularly if the disk is diamagnetic. This steepening of the field results
in a flattening of the $\nuqpo-\mdot$ relation. Additional flattening
results from incipient general relativistic instability at the inner
edge of the disk. 

\subsection{A Specific Ansatz: Diamagnetic Disk}

An illustration of the field steepening discussed above
is as follows. Consider a vacuum dipole field produced by the star
$|B_z|=\mu/r^3$ (in the equatorial plane perpendicular to the dipole axis).
Imagine inserting a diamagnetic disk 
in the equatorial plane with inner radius $r_{\rm in}$. Flux conservation
requires $\pi (r_{\rm in}^2-R^2)|{\bar B_z}|=2\pi \mu/R$, which gives
the mean vertical field inside between $R$ and $r_{\rm in}$: 
\be
|\bar B_z(r_{\rm in})|={2\mu\over R(r_{\rm in}^2-R^2)}.
\label{eq:Baver}\ee
This field has scaling $|\bar B_z|\propto 1/r^2$ for large $r$, which would 
result in $\nu_K(r_{\rm in})\propto \dot M$ (see
eq.~[\ref{eq:scale}]), and stiffens as the disk approaches the stellar
surface. 

The actual field at $r=r_{\rm in}$ is difficult to calculate. Aly (1980) 
found the magnetic field of a point dipole in the presence of
a thin diamagnetic disk (thickness $H\ll r$ at radius $r$), 
and demonstrated that the field strength
at $r_{\rm in}$ is enhanced by a factor
$\sim (r_{\rm in}/H)^{1/2}$.
(See also Riffert 1980 and Arons 1993.)
However, the situation is different for a 
finite-sized dipole (a conducting sphere of radius $R$) in the presence of 
a diamagnetic disk. This can be seen by 
considering a simpler problem, where we replace the disk
by a diamagnetic sphere (with radius $r_{\rm in}$)\footnote{ In replacing the
disk with a spherical surface, we lose the square-root divergence found by
Aly (1980) for infinitesmal $H/r$. But note that for small fields, the
disk penetrates near the star, and $H$ may not be {\it very} small compared
with $\rin-R$. In assuming a point dipole, Aly (1980) (and Riffert 1980)
exacerbated the divergence, and their results probably apply only when
$\rin\gg R$.}.The magnetic field at
radius $r$ (between $R$ and $r_{\rm in}$) is given by (in spherical
coordinates with the magnetic dipole along the $z$-axis):
\begin{eqnarray}
B_r(r,\theta) &=& 
\left(-{2\mu\over r_{\rm in}^3}+{2\mu\over r^3}\right){\cos\theta\over
1-\alpha^3},\\
B_\theta (r,\theta) &=&
\left({2\mu\over r_{\rm in}^3}+{\mu\over r^3}\right){\sin\theta\over
1-\alpha^3},
\end{eqnarray}
where $\alpha=R/r_{\rm in}$. Thus the vertical magnetic field at the inner edge
of the disk ($r=r_{\rm in}$) is
\be
|B_z(r_{\rm in})|={3\mu\over r_{\rm in}^3-R^3}.
\label{eq:moddipole}\ee
We see that the magnetic field steepens as $r_{\rm in}$ approaches the stellar
surface. In reality, some magnetic field will penetrate the disk because of
turbulence in the disk and Rayleigh-Taylor instabilities (Kaisig, Tajima
\& Lovelace 1992); however, some steepening of the field may remain.

Adopting the magnetic field {\it ansatz} (\ref{eq:Baver}) and
$B_\phi=-\beta B_z$, we can use (\ref{eq:balance})
to calculate $r_{\rm in}$; this gives
\be
2\,b^2{x_c^{2.5}\over (x_c^2-1)^2}=1,
\label{eq:rinone}
\ee
where $x_c=r_{\rm in}/R$ and
\be
b^2={\beta B_0^2R^3\over\dot M\sqrt{GMR}}=
0.07\left(M_{1.4}^{-1/2}R_{10}^{5/2}\right)
\left({\beta B_7^2\over \dot M_{17}}\right),
\ee
$\mu=B_0^2R^3/2$ ($B_0=10^7B_7$~G is the polar 
field strength at the neutron star surface),
$M_{1.4}=M/(1.4\,M_\odot)$, $R_{10}=R/(10\,{\rm km})$, 
and $\dot M_{17}=\dot M/(10^{17}\,{\rm g\,s}^{-1})$.
Alternatively, if we adopt (\ref{eq:moddipole}), we find
\be
{9\over 2}\,b^2{x_c^{2.5}\over (x_c^3-1)^2}=1.
\label{eq:rintwo}
\ee
Figure 1 shows the Kepler frequency at $r_{\rm in}$ as a function the
scaled mass accretion rate, $\mdot_{17}M_{1.4}^{1/2}R_{10}^{-5/2}
M_{17}/\beta B_7^2=0.07/b^2$. Clearly, for small $\dot M$, $\nu_K(r_{\rm
in})$ depends on $\dot M$ through a power-law, but the dependence weakens as 
$\dot M$ becomes large, in qualitative agreement with the 
observed $\nu_{\rm QPO}$-$\dot M$ correlation.
General relativistic effects also flatten 
the $\nu_{\rm QPO}$-$\dot M$ relation, as we discuss next.

\subsection{General Relativistic Effects}

General relativity (GR) introduces two effects on the location of the
inner edge of the disk. First, the space-time curvature modifies 
the vacuum dipole field. For example, in Schwarzschild metric, the locally
measured magnetic field in the equatorial plane is given by
\be
B^{\hat\theta}={\mu\over r^3}\left[6y^3(1-y^{-1})^{1/2}\ln(1-y^{-1})
+{6y^2(1-y^{-1}/2)\over (1-y^{-1})^{1/2}}\right],
\ee
where $y=rc^2/(2GM)$ (Petterson 1974; Wasserman \& Shapiro 1983).
The GR effect steepens the field only at small $r$. For $r=6GM/c^2-10GM/c^2$, 
we find the approximate scaling $B^{\hat\theta}\propto r^{-3-\epsilon}$, 
with $\epsilon \simeq 0.3-0.4$. We shall neglect such a small
correction to the dipole field given the much larger uncertainties
associated with the magnetic field -- disk interaction.

A more important effect of GR is that it modifies the the dynamics of
the accreting gas around the neutron star. 
Without magnetic field, the inner edge of the disk
is given by the condition $d l_K/dr=0$, where $l_K$ is the
specific angular momentum of a test mass:
\be
l_K=\left({GMr^2\over r-3GM/c^2}\right)^{1/2}.
\ee
This would give the usual the ISO at $r_{\rm iso}=6GM/c^2$,
where no viscosity is necessary to induce accretion\footnote{When viscosity
and radial pressure force is taken into account, the flow is transonic,
with the sonic point located close to $r_{\rm iso}$.}. 
Since magnetic fields take angular momentum
out of the disk, we can determine the inner edge of the disk using an
analogous expression\footnote{Note that in the limit of perfect conductivity,
it is possible to express the Maxwell stress tensor in terms of a 
magnetic field four-vector
$\bf B$ that is orthogonal to the fluid velocity four-vector
$\bf U$ (e.g. Novikov \& Thorne 1973, pp. 366-367). The field 
components $B_\phi$ and $B_z$
in eq. (\ref{eq:balance2}) and below are actually the projections
of $\bf B$ onto a local orthonormal basis (i.e. $B_\phi\to\vec{\bf e}_{\hat\phi}
\cdot{\bf B}$ and $B_z\to\vec{\bf e}_{\hat z}\cdot{\bf B}$)
even though we have retained the nonrelativistic notation for these
field components. No additional relativistic corrections are required with
these identifications understood.}
\be
\dot M {dl_K\over dr}=-r^2B_zB_\phi,
\label{eq:balance2}\ee
(see eq.~[\ref{eq:balance}]). In Lai (1998) it was shown that 
this equation determines the limiting value of the sonic point 
of the accretion flow (although a Newtonian pseudo-potential was
used in that paper). Adopting the magnetic field {\it ansatz} (\ref{eq:Baver}), 
we find
\be
2\,b^2{x_c^{2.5}\over (x_c^2-1)^2}=\left(1-{6GM\over c^2r_{\rm in}}\right)
\left(1-{3GM\over c^2r_{\rm in}}\right)^{-3/2}.
\label{eq:gr1}\ee
Similarly, using (\ref{eq:moddipole}), we have
\be
{9\over 2}\,b^2{x_c^{2.5}\over (x_c^3-1)^2}=\left(1-{6GM\over c^2r_{\rm
in}}\right)\left(1-{3GM\over c^2r_{\rm in}}\right)^{-3/2}.
\label{eq:gr2}\ee
It is clear that for $b\gg 1$, eq.~(\ref{eq:gr1}) or (\ref{eq:gr2})
reduces to the Newtonian limit (see \S 3.1), while for $b=0$ we recover
the expected $r_{\rm in}=r_{\rm iso}=6GM/c^2$. For small $b$, the GR
effect can modify the inner disk radius signficantly.
In Fig.~1 we show the orbital frequency at $r_{\rm in}$ 
(as a function of the ``effective'' accretion rate) as obtained 
from (\ref{eq:gr1}) and (\ref{eq:gr2}). We see that the GR effect induces additional
flattening in the correlation between $\nu_K(r_{\rm in})$ and $\dot M$ 
as $r_{\rm in}$ approaches $r_{\rm iso}$. 

We emphasize the phenomenological nature of
eqs.~(\ref{eq:balance2})-(\ref{eq:gr2}): they are not derived from a 
self-consistent MHD calculation, and take account of the dynamics 
of the disk under a prescribed magnetic field configuration. 
However, we believe that they indicate the combined effects
of dynamically altered magnetic field 
and GR on the inner region of the accretion disk. 
By measuring the correlation between the QPO frequency and the mass 
accretion rate, one might be able to
constrain the magnetic field structure in accreting 
neutron stars, and reach quantitative conclusions about
the nature of the interaction of the accretion disk and
magnetic field.


\section{Where are the QPOs Produced?}

Implicit in the discussion of magnetic fields and $\nuqpo$ in the
preceding sections were the assumptions that the QPO arises at
a radius outside the star that coincides with the inner radius
of the accretion disk. Here, we examine two ways
in which these assumptions might be violated, and show how the
relatively small measured values of $\numax$ might be consistent
with neutron star masses near $1.4\msun$.

\subsection{Disk Termination at the Neutron Star Surface}

For the model discussed in \S 3, the steepening magnetic field
and general relativity produce the flattening in the correction between
the QPO frequency $\nu_{\rm QPO}=\nu_K(r_{\rm in})$ and the mass 
accretion rate $\dot M$. But $\nu_{\rm QPO}$ becomes
truly independent of $\dot M$ only when $r_{\rm in}$ approaches
$r_{\rm iso}$ or the stellar radius $R$. It has been suggested (see \S 1)
that the $\dot M$-independent QPO frequency corresponds to 
the Kepler frequency at $r_{\rm iso}$. But it is also possible
that the inner disk radius reaches the stellar surface, which is outside
the ISO,
as $\dot M$ increases. We note that observationally it is difficult 
to distinguish the flattening of $\nu_{\rm QPO}$ and a true plateau. 
It is not clear that the flattening feature at $\nu_{\rm QPO}\sim 1100$~Hz
observed in 4U 1820-30 (Zhang et al.~1998) corresponds the maximum 
QPO frequency, but we shall assume it does and explore the consequences. 


The maximum QPO frequency, $\nu_{\rm max}$,
is given by the orbital frequency at the larger of $r_{\rm iso}$ and $R$.
To linear order in $\nu_s$ (the spin frequency), the ISO is located at 
$r_{\rm iso}=(6GM/c^2)(1-0.544\,a)$, and the orbital frequency at ISO is 
\be
\nu_K(r_{\rm iso})={1571\over M_{1.4}}(1+0.748\,a)~{\rm Hz},
\ee
with the dimensionless spin parameter 
\be
a\simeq 0.099\, {R_{10}^2\over M_{1.4}}\left({\nu_s\over 300\,{\rm Hz}}
\right),
\ee
where we have adopted $I=(2/5)\kappa MR^2$ for the moment of inertia of 
the neutron star, with $\kappa\simeq 0.815$ (appropriate for a 
$n=0.5$ polytrope). The orbital frequency at the stellar surface can be
written, to linear order in $\nu_s$, as
\be
\nu_K(R)=2169\,M_{1.4}^{1/2}R_{10}^{-3/2}\left[1-0.094\,a\,
\left({M_{1.4}\over R_{10}}\right)\right]\,{\rm Hz}.
\ee
Note that in the above equations, $R$ refers to the equatorial radius
of the (spinning) neutron star, which is related to the radius, $R_0$,
of the corresponding nonrotating star by: 
\be
{R-R_0\over R_0}\simeq 0.4{\Omega_s^2 R_0^3\over GM}
\simeq 0.0078\,M_{1.4}^{-1}R_{10}^3\left({\nu_s\over 300\,{\rm Hz}}
\right)^2,
\ee
where we have again adopted the numerical parameters 
appropriate for a $n=0.5$ polytrope (Lai et al.~1994). 
One may appeal to numerical calculations (e.g., Miller, Lamb \& Cook
1998) for more accurate results, but the approximate expressions given above
are adequate. 

Figure 2 shows the contours of constant $\nu_{\rm max}={\rm min}[\nu_K(r_{\rm
iso}),\nu_K(R)]$ in the $M$-$R_0$ plane. For large $M$ and small $R_0$,
the contours are specified by $\nu_K(r_{\rm iso})$, while for larger $R_0$
and small $M$, the contours are specified by $\nu_K(R)$. 
We see that to obtain the maximum 
QPO frequency of order $1100-1200$~Hz, one can either have 
a $M\go 2M_\odot$ neutron star (with $R_0\lo 16$~Km), or have 
a $M\simeq 1.4M_\odot$ neutron star with $R_0\simeq 14-15$~km. Here we  
focus on the latter interpretation, in which 
the accretion disk
terminates at the stellar surface before reaching the ISO. A 
boundary layer forms in which the angular velocity of the accreting
gas changes from near the Keplerian value (at the outer edge of the
boundary layer) to the stellar rotation rate. Depending on the thickness
of the boundary layer, the inferred the NS radius may be somewhat 
smaller.  Moreover, the peak rotation frequency may be below $\nuk(R)$,
which would also allow smaller values of $R_0$.

In addition to avoiding a large neutron star mass (see \S2), 
the identification of $\nu_{\rm max}$ with
the Kepler frequency near the stellar surface may allow 
a plausible explanation of the observed correlation between
the QPO amplitude and the X-ray flux. While the mechanism 
of producing X-ray modulation in a kHz QPO is uncertain,
in many models (e.g., Miller et al.~1998; see also Klu\'zniak et al.~1990)
the existence of a supersonic ``accretion gap'' between the
stellar surface and the accretion disk is crucial for
generating the observed the X-ray modulation. 
If we interperate $\numax$ as
the Kepler frequency at the ISO, which is always outside the stellar
surface, then the ``accretion gap'' always exists, 
and there is no qualitative change in the flow behavior as the inner disk
approaches ISO. It is therefore difficult to explain why the QPO
amplitude decreases and eventually vanishes as the X-ray flux increases.
The situation is different if $\numax=\nuk(R)$, since the gap disappears
when the mass accretion rate becomes sufficiently large.
At small $\dot M$ there is a gap (induced by a combination
of magnetic and GR effects) between the inner edge 
of the disk and the stellar surface. Since the impact velocity 
of the gas blob at the stellar surface is larger for a wider accretion gap,
we expect the modulation amplitude to be larger for small accretion rates
\footnote{When $\dot M$ is too low (for a given $B_0$)
so that $r_{\rm in}$ is far away from the 
stellar surface, the accreting gas can be channeled out of the disk plane by
the magnetic field toward the magnetic poles. The detail
of the channeling process depends on the magnetic field geometry  
in the disk (such as the radial pitch angle of the field line). 
This may quench the kHz QPOs
and give rise to X-ray pulsation (as in X-ray pulsars).
The pulsating X-ray transient system SAX J1808.4-3658 may just be such an
example.}.
As $\dot M$ increases, the inner disk edge approaches the stellar surface,
and we expect the QPO amplitude to decrease. The maximum QPO 
frequency signifies the closing of the accretion gap and the formation of a
boundary layer. Since there is no supersonic flow in this case, 
one might expect the QPO amplitude to vanish.  In addition, there may be
changes in the spectral properties of the system as the gap closes.

The large neutron star radius ($15$~km for a $1.4M_\odot$ star)
required if $\numax=\nu_K(R)$ is only allowed
for a handful of very stiff
nuclear equations of state (see Fig.~2); most recent microscopic 
calculations give $R_0\sim 10$~km (e.g., Wiringa et al.~1988). 
Is such a large radius consistent with observations? 
No neutron star radii are known with the accuracy that has been
achieved for numerous neutron star mass determinations, but several 
methods have been tried:
\begin{enumerate}

\item Observations of X-ray bursts have been
used to determine empirical $M-R$ relations, but these are
hampered by the need for model-dependent assumptions regarding the total
luminosity and its time history, anisotropy of the emission, radiated
spectrum and surface composition, even when the source distance is
known (e.g. van Paradijs et al. 1990, Lewin, van Paradijs \& Taam 1995).

\item X-ray
and optical observations of the (apparently nonrotating) isolated
neutron star RX J185635-3754 (Walter, Wolk \& Neuh\"auser 1996, Walter \& Matthews
1997), combined with limits on the source distance, $D$, imply a
blackbody radius $R(1+z)<14(D/130\,{\rm pc})$ km, where $z$ is the
surface redshift of the star. 

\item Ray tracing and lightbending may be used to derive limits
on $R/M$ for periodically modulated X-ray emission. For two isolated
neutron stars (PSR B1929+10 and B0950+08; Yancopoulos, Hamilton \& Helfand 1994,
Wang \& Halpern 1997) and one millisecond pulsar (J0437-4715; Zavlin
and Pavlov 1997, Pavlov \& Zavlin 1998), the results are broadly consistent with
$Rc^2/2GM\simeq 2.0-2.5$, but the results depend on geometry (angles
between rotation and magnetic axes, and rotation axis and the line of
sight) as well as on the spectrum and (energy-dependent) anisotropy
of the polar cap emission. The rather large observed pulsed fractions
appear to rule out two polar cap hot spots unless $Rc^2/2GM$ is
rather large (e.g. $\simeq 4.3$ for PSR B1929+10; Wang \& Halpern
1997). Similar considerations may prove fruitful for periodically
modulated flux from X-ray bursts (e.g. Miller \& Lamb 1998); the
pulse fractions observed so far are large, suggesting non-compact
sources (e.g. Strohmayer et al. 1999, who find $Rc^2/2GM\simeq 5$
for 4U1636-54, corresponding to an implausibly large
radius of 21 km for $M=1.4\msun$). 

\item Burderi \& King (1998) have argued that requiring the Alfv\'en
radius to be intermediate between $R$ and the corotation radius,
$R_{co}=(GM/\Omega_s^2)^{1/3}$, for the 2.5 ms
pulsating source SAX J1808.4-3658 (discovered by Wijnands \& van der
Klis 1998)
implies an upper bound of $R<13.8(M/\msun)^{1/3}$ km, since the pulsations
are detected at the same frequency for X-ray count rates spanning an
order of magnitude. However, their bound
depends on the model-dependent assumptions
that the count rate is strictly proportional to $\mdot$ and the field
strength in the disk is dipolar ($B\propto r^{-3}$).
(See also Psaltis \& Chakrabarty 1999.)

\end{enumerate}

\noindent Taken together, the evidence neither supports nor excludes the
possibility that $R\simeq 15$ km for $M\simeq 1.4\msun$ (or $Rc^2/2GM
\simeq 3.6$) definitively, although most of the estimates listed above
favor more compact models ($R\simeq 10$ km for $M\simeq 1.4\msun$)
nominally.

\subsection{QPOs from $r>\rin$?}

QPOs are identified in the Fourier spectra of photon counts from X-ray
sources, so it may be that most of the spectral power comes from radii
outside $\rin$, possibly from the disk radius at which the differential
photon emission rate is maximum. For example, if the QPO arises from
a radius $r=(1+\lambda)\rin$, then $\nuqpo=(1+\lambda)^{-3/2}
\nuk(\rin)$. As $\rin\to 6GM/c^2$, the ISO in the slow-rotation limit,
$\nuqpo\to 2200\,\hz/(M/\msun)(1+\lambda)^{3/2}$, so observations that
give $\numax\simeq 1060\,\hz$ asymptotically may actually require
$M(1+\lambda)^{3/2}=2.1\msun$, or $1+\lambda\approx 1.3$ if 
$M\approx 1.4\msun$. rather than $M\simeq 2.1\msun$.

To obtain a simple realization of this idea, consider a Shakura-Sunyaev
(1973) disk, for which the emitted flux from one face is
\be
F(r)=\sigma_{\rm SB}T_e^4(r)={3GM\mdot f(r)\over 8\pi r^3};
\ee
in the Newtonian limit (which we shall employ here for giving a
simplified illustration). The function
$f(r)=1-\beta\sqrt{\rin/r}$, where
$\beta\leq 1$ parametrizes the rate of accretion of
angular momentum from the disk onto the star relative to 
$\mdot\sqrt{GM\rin}$
(e.g. Shapiro \& Teukolsky 1983. eq. [14.5.17]; see also
Frank, King \& Raine 1992, \S 5.3); if ``imperfect'' fluid stresses vanish
at $\rin$, then $\beta=1$ (as in black hole accretion; see Page
\& Thorne 1974, Novikov \& Thorne 1973). If the
color temperature of the emission equals the effective temperature
$T_e(r)$, then the ``bolometric flux'' of photons is $\sim F(r)/kT_e(r)$
at radius $r$, and the rate at which photons are emitted from radii
between $r$ and $r+dr$ is of order
\be
{2\pi rF(r)\over kT_e(r)}\sim {\mdot^{3/4}[f(r)]^{3/4}\over
r^{5/4}}.
\label{eq:noemit}
\ee
Differentiating equation (\ref{eq:noemit}) implies a maximum emission
rate at $\sqrt{r/\rin}=1.3\beta$, consistent with $r>\rin$ provided that
$\beta>0.77$. Assuming that the QPO frequency is the Kepler frequency
at the radius of peak (bolometric) photon emission,
\be
\nuqpo = {\nuk(\rin)\over (1.3\beta)^3}
\to{1000\,\hz\over\beta^3M/\msun},
\ee
where the limiting result
is for $\rin\to 6GM/c^2$. In order for the maximum
value of $\nuqpo$ to be $\numax\simeq 1060$ Hz, we require
$M=1.4\msun/(\beta/0.88)^3$.

Real disk emission profiles for small $\rin$, and the determination of
$\nuqpo$, are not this simple for several reasons. A detailed
calculation of the X-ray spectrum is needed, since the QPOs are found
for counts in particular energy bands; the bolometric count rate is
not a good approximation in general. (But note that Comptonization by
hot coronal gas above the disk conserves photon number, so the
approximation may be better than it appears at first sight.) In
particular, the color temperature is not usually the same as the 
effective temperature, since electron scattering is the dominant
opacity at relevant disk radii. The composition of the disk is also
important; at low enough $\mdot$, the disk will be matter-dominated, but
at larger $\mdot$, radiation-dominated. (Less important, but still
significant, is the dependence of opacity on the element abundances
in the accreting gas.) In addition, relativistic effects alter
$f(r)$ (e.g. Page \& Thorne 1974, Novikov \& Thorne 1973), and hence
$\nuqpo$. Moreover, the angular momentum carried away by photons
may not be insignificant once $\rin$ approaches the ISO (Page \&
Thorne 1974, Epstein 1985). Instabilities associated with the transition
from matter to radiation domination (Lightman \& Eardley 1974) or the
inner boundary layer (e.g. Epstein 1985) might also play a role in 
determining $\nuqpo$. These and other issues associated with the 
termination of disks at $\rin$ and QPOs will be explored more fully
elsewhere. However, the simplified example presented here 
indicates that $\nuqpo$ might plausibly arise from 
$r>\rin$.


\section{Conclusion}

In this paper we have presented a phenomenological model 
of the inner region of the accretion disk for weakly magnetized 
neutron stars such as those in LMXBs. A notable feature of these systems
is that both magnetic field and general relativity are important in
determining the inner disk radius. Our result suggests that the combined
effects of a steepening magnetic field -- which is likely for disk
accretion onto a neutron star --  
and general relativity can produce the flattening of
the QPO frequency $\nu_{\rm QPO}$ as the mass accretion rate $\dot M$ 
increases. If the field steepens fast enough with decreasing inner
disk radius, $\nuqpo$ may vary little over a fairly substantial range
of $\mdot$ at values considerably below the
Kepler frequency at the ISO due to general relativity.
Observationally, the correlation between $\nuqpo$ and 
the RXTE photon count rate has been well-established, but the scaling 
between $\nu_{\rm QPO}$ and $\dot M$ is ambiguous (Mendez et al.~1998c). 
An observational or phenomenological
determination of this scaling would be quite useful 
in constraining the magnetic field structure in LMXBs.

Currently it is not clear whether the plateau behavior in 
the QPO frequency has been observed. But even if $\nu_{\rm QPO}\sim 1100
-1200$~Hz represents the maximum possible QPO frequency, we argue that 
a massive neutron star ($M\go 2M_\odot$) is not necessaily implied. 
Instead, a $M\simeq 1.4M_\odot$, $R_0\sim 14-15$~km neutron star
may be a better solution, and is within the range allowed by
some nuclear equations of state. If this is the case, 
the maximum QPO frequency signifies the closing of the accretion gap and the
formation of a boundary layer. 
Alternatively, the QPO frequency might be associated with the Kepler
frequency at a radius somewhat larger than the inner radius of the disk,
thus allowing lower mass for the accreting neutron star. In either case,
better theoretical and phenomenological
understanding of the termination of magnetized accretion disks
is needed before observations of maximal kHz QPOs can be interpreted as purely
general relativistic in origin, and used to deduce neutron star masses.

\acknowledgments
D.L. is supported
by a Alfred P. Sloan Foundation Fellowship.
R.L. acknowledges support from NASA grant NAG 5-6311.
I.W. acknowledges support from NASA grants NAG 5-3097 and NAG 5-2762.


\bigskip

\begin{figure}
\plotone{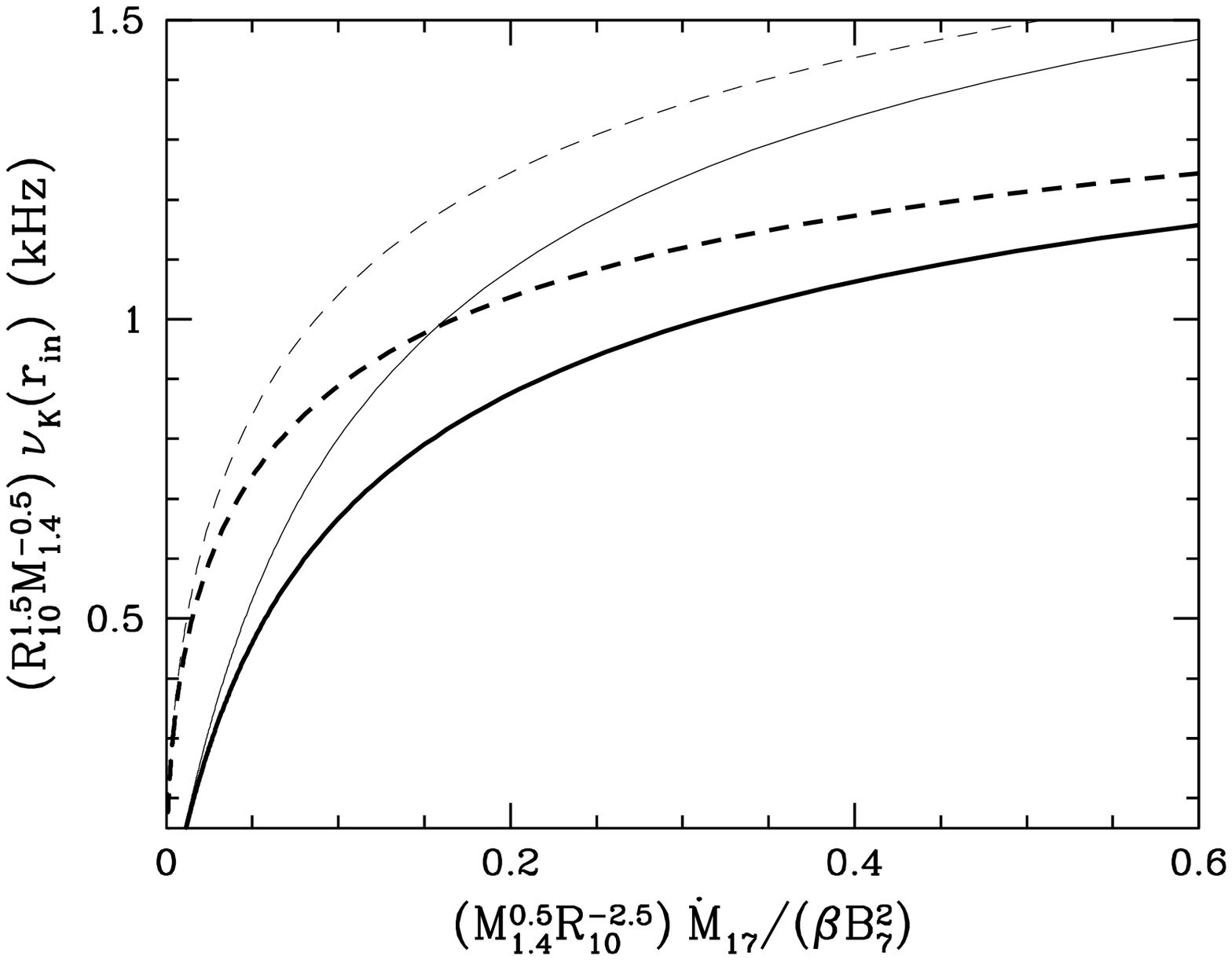}
\caption{
The orbital frequency of at the inner radius of the the disk
as a function of the ``effective'' mass accretion rate. 
The inner disk radius, $r_{\rm in}$, is obtained by solving
eq.~(\ref{eq:rinone})
(light solid line) or (\ref{eq:rintwo}) (dashed solid line), corresponding to
different magnetic field structure. The heavy solid lines
incorporate the effect of general relativity based on eq.~(\ref{eq:gr1})
(heavy solid line) or (\ref{eq:gr2}) (heavy dashed line). 
\label{fig1}}
\end{figure}

\begin{figure}
\plotone{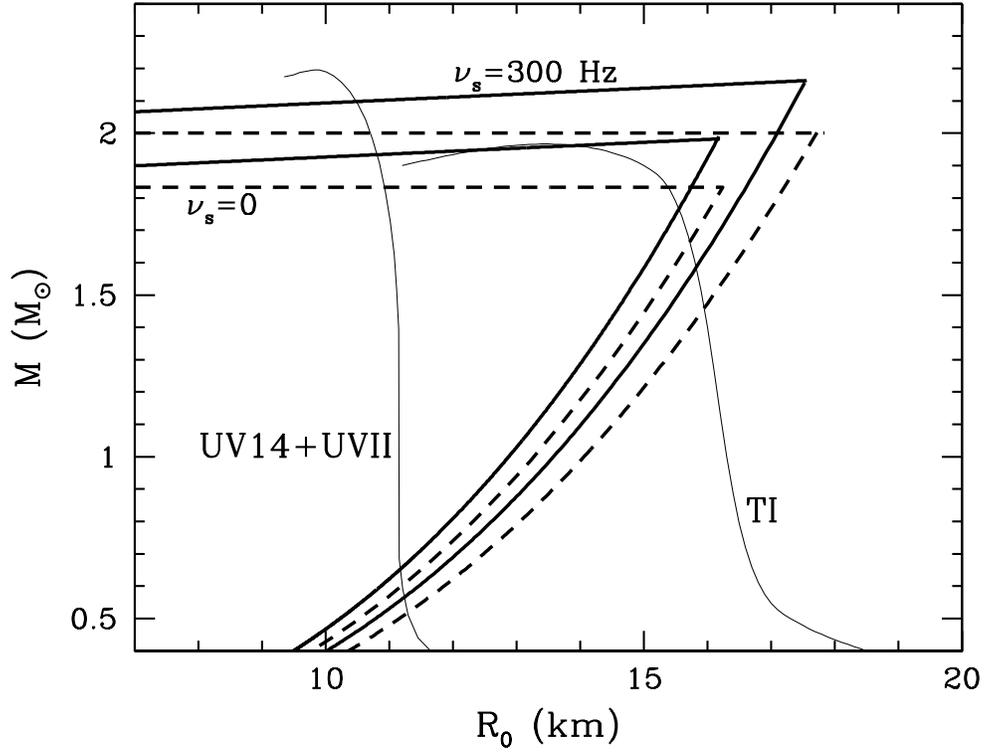}
\caption{
Constraints on the mass-radius ($M-R_0$) relation of neutron star from the
maximum orbital frequency $\nu_{\rm max}$ outside the star. 
Each closed curve shows the $\nu_{\rm max}=$constant contour in the $M-R_0$
plane, with the upper boundary 
$\nu_{\rm max}=1100$~Hz, and the lower boundary $\nu_{\rm max}=1200$~Hz.
(the solid heavy lines correspond to spin frequency $\nu_s=300$~Hz, and
the heavy dashed lines $\nu_s=0$). 
Note that in the case of rotating neutron star, 
$R_0$ is the radius of corresponding nonrotating stellar model with the
same mass. The light solid curves depict two representative equations of 
state: TI is the very stiff tensor interaction model of 
Pandharipande \& Smith (1975), and UV14+UVII is from model of 
Wiringa et al.~(1988).
\label{fig2}}
\end{figure}

\end{document}